# Identifying Cancer Patients at Risk for Heart Failure Using Machine Learning Methods

Xi Yang, PhD[1], Yan Gong, PhD, MS[2,3], Nida Waheed, MD[4], Keith March, MD, PhD[5], Jiang Bian, PhD[1], William R. Hogan, MD, MS[1], Yonghui Wu, PhD[1]

Department of [1]Health Outcomes and Biomedical Informatics, [2]Pharmacotherapy and Translational Research, [3]Center for Pharmacogenomics and Precision Medicine, [4]Internal Medicine, and [5]Division of Cardiovascular Medicine, College of Medicine, University of Florida, Gainesville, Florida, USA

**Abstract**

*Cardiotoxicity related to cancer therapies has become a serious issue, diminishing cancer treatment outcomes and quality of life. Early detection of cancer patients at risk for cardiotoxicity before cardiotoxic treatments and providing preventive measures are potential solutions to improve cancer patients' quality of life. This study focuses on predicting the development of heart failure in cancer patients after cancer diagnoses using historical electronic health record (EHR) data. We examined four machine learning algorithms using 143,199 cancer patients from the University of Florida Health (UF Health) Integrated Data Repository (IDR). We identified a total number of 1,958 qualified cases and matched them to 15,488 controls by gender, age, race, and major cancer type. Two feature encoding strategies were compared to encode variables as machine learning features. The gradient boosting (GB) based model achieved the best AUC score of 0.9077 (with a sensitivity of 0.8520 and a specificity of 0.8138), outperforming other machine learning methods. We also looked into the subgroup of cancer patients with exposure to chemotherapy drugs and observed a lower specificity score (0.7089). The experimental results show that machine learning methods are able to capture clinical factors that are known to be associated with heart failure and that it is feasible to use machine learning methods to identify cancer patients at risk for cancer therapy-related heart failure.*

**Introduction**

Cancer is the second leading cause of death in the US.[1] There has been a great amount of effort and resources invested in the development of new cancer therapies. The mortality rates of many cancers are being brought under control with the improvement of cancer treatment.[2] However, these anticancer treatments often have various side effects. For example, cardiotoxicity is one of the well-documented adverse events of cancer treatments resulting either from accelerated development of cardiovascular diseases in cancer patients or from the direct effects of the treatment on the structure and function of the heart.[3] Traditional chemotherapy such as anthracyclines have been known to cause cardiovascular complications.[4–6] Cardiotoxicity related to cancer therapies has become a serious issue that diminishes cancer treatment outcomes. A recent study examined various anticancer therapies and reported a significant correlation between quality of life (QoL) and chemotherapy cycles.[7] Early detection and possible prevention of cardiotoxicity in cancer treatments is a potential solution to improve cancer patients' safety and QoL. Identifying cancer patients with high risk of cardiotoxicity is a critical step towards early detection and possible prevention.

In the last two decades, the introduction of targeted anticancer therapies has revolutionized the treatment of both hematological malignancies such as multiple myeloma, chronic myeloid leukemia and solid malignancies such as breast and renal carcinoma.[8,9] Contemporary cancer therapy has led to a 23% reduction in cancer-related mortality rate and rapid increase in cancer survivorship in the last 15 years.[10] However, some devastating side effects of these treatments have also resulted in increased morbidity and mortality.[11,12] Examples of these targeted cancer therapies include human epidermal growth factor 2 inhibitors, inhibitors of vascular endothelial growth factor pathway and tyrosine kinase inhibitors and proteasome inhibitors. Most recently, immune checkpoint inhibitors have also been associated with cardiotoxicity.[13,14] Despite the efficacy of these therapies, their widespread use has paradoxically resulted in the emergence of serious cardiovascular effects/complications such as cardiomyopathy/heart failure, coronary artery disease, myocardial ischemia, hypertension, arrhythmia, thromboembolism, and pericardial disease.[15] One of the most relevant clinical implications of these complications is treatment interruption, which is associated with cancer recurrence. Due to the high incidence and negative impact on patient outcomes, new medical subspecialties such as Cardio-Oncology were created to optimize the care or management of patients receiving these cancer therapies. Identifying patients with high risk of cardiotoxicity using historical electronic health records (EHRs) could be potentially used to improve cancer treatment safety and QoL.

Rapid adoption of EHRs has made longitudinal clinical data available to research. There is an increasing interest in using longitudinal EHRs to develop computational algorithms for disease onsite prediction. Researchers have applied standard statistical regression models and machine learning methods to predict the onsite of heart failure among general patient cohorts. For example, Wang et al. developed a heart failure predicting model using random forests (RFs) and examined various prediction windows[16]; Sun et al. proposed a method to combine knowledge and data driven method to identify risk factors of heart failure from EHRs[17]; Wu et al. compared three machine learning models including Boosting, support vector machines (SVMs) and logistic regression (LR) for heart failure prediction.[18] While machine learning-based predictive models showed decent performance, previous studies identified issues such as imbalanced data[18] and the lack of modeling temporal sequence among clinical events. Recently, Choi et al. applied recurrent neural networks (RNNs) for heart failure prediction and compared RNN with a traditional machine learning model – SVMs.[19] Their study reported that deep learning models were able to leverage temporal relations among clinical events to improve performance of heart failure prediction with a short observation window of 12-18 months. Rasmy et al. also examined the generalizability of RNN in predicting heart failure onset risk using a large and heterogeneous EHR data set. Researchers from the cancer community have applied statistical regression models for risk assessment of heart failure after cancer treatments. For example, Ezaz et al. applied regression analysis to assess risk scores of heart failure among breast cancer patients after trastuzumab therapy.[20] Authors used a study cohort consisted of women from 67 to 94 years old from the SEER-Medicare database diagnosed with early-stage breast cancer. Although machine learning has been successfully applied to the general patient cohorts for heart failure prediction, it's not clear whether it can be applied to cancer patient cohorts to support early detection and possible prevention of cardiotoxicity in cardiotoxic cancer therapies.

This study focused on the prediction of cancer patients developing heart failure after cancer diagnoses. We examined four machine learning algorithms for heart failure prediction among cancer patients from the University of Florida Health (UF Health) Integrated Data Repository (IDR). We compared four widely used machine learning models including LR, RFs, SVMs, and Gradient Boosting (GB) for heart failure prediction among cancer patients. We systematically examined variables including patient demographics (gender, race, age), diagnoses, medications, and procedures as machine learning features. We also compared one-hot encoding and term frequency-inverse document frequency (TF-IDF) encoding for all four machine learning methods. Furthermore, we compared the performance of heart failure prediction on general cancer patients with the performance on cancer patients with exposure to chemotherapy drugs. Our ultimate goal is to develop predictive models for identification of cancer patients with high risk of cardiotoxicity to prevent or minimize the risk of cardiotoxicity in cancer treatments.

**Methods**

**Data set**

In this study, we used EHR data from UF Health IDR. Supported by the UF Clinical and Translational Institute (CTSI) and the UF Health, the UF Health IDR is a secure, clinical data warehouse (CDW) that aggregates data from the university's various clinical and administrative information systems, including the Epic electronic medical record (EMR) system. As of February 2019, the IDR contains data for encounters that occurred after June 2011, with a total of more than 1105 million observational facts pertaining to 1.17 million patients. From UF Health IDR, we collected a total number of 143,199 cancer patients from Jan 1st, 2011 to Dec 31st, 2017. We identified cancer patients using the diagnosis codes (at least one cancer diagnosis code). The extracted EHR data contains patient demographics, diagnoses in both the International Classification of Disease version 9 (ICD-9) and the International Classification of Disease version 10 (ICD-10), medications in RXNORM codes, and procedures in CPT codes. This study was approved by the UF Institutional Review Board.

**Definition of cases and controls**

Starting from the initial cancer patients, we removed patients diagnosed with only benign type of cancers. Then, we identified the first diagnosis date of cancer as cancer index date, denoted as CAID. Following previous studies from Choi et al.[19] and Rasmy et al.[21], we defined the heart failure onset date (HFOD) as the first encounter date of three consecutive heart failure diagnosis encounters occurred within 12 months.

*Cases*: Following previous studies on heart failure prediction of general patient cohorts, cases of heart failure were defined as follows: (1) having at least three encounters with qualified heart failure diagnoses defined in Table 1 occurred within 12 months; (2) the CAID must before the HFOD – this rule removed the cancer patients with existing heart failure conditions before cancer diagnosis; (3) at least have one medication record or one procedure record – we want to ensure that the patient received treatments at UF Health.

*Controls*: Controls were defined as cancer patients without any qualified heart failure codes defined in Table 1.

**Table 1.** ICD-9 and ICD-10 codes used for heart failure diagnosis.

| ICD-9 | Description | ICD-10 | Description |
| --- | --- | --- | --- |
| 428 | heart failure | I50 | heart failure |
| 428.0 | congestive heart failure, unspecified | I50.1 | left ventricular failure, unspecified |
| 428.1 | left heart failure | I50.2 | systolic (congestive) heart failure |
| 428.2 | systolic heart failure | I50.20 | unspecified systolic (congestive) heart failure |
| 428.20 | systolic heart failure, unspecified | I50.21 | acute systolic (congestive) heart failure |
| 428.21 | acute systolic heart failure | I50.22 | chronic systolic (congestive) heart failure |
| 428.22 | chronic systolic heart failure | I50.23 | acute on chronic systolic (congestive) heart failure |
| 428.23 | acute on chronic systolic heart failure | I50.3 | diastolic (congestive) heart failure |
| 428.3 | diastolic heart failure | I50.30 | unspecified diastolic (congestive) heart failure |
| 428.30 | diastolic heart failure, unspecified | I50.31 | acute diastolic (congestive) heart failure |
| 428.31 | acute diastolic heart failure | I50.32 | chronic diastolic (congestive) heart failure |
| 428.32 | chronic diastolic heart failure | I50.33 | acute on chronic diastolic (congestive) heart failure |
| 428.33 | acute on chronic diastolic heart failure | I50.4 | combined systolic (congestive) and diastolic (congestive) heart failure |
| 428.4 | combined systolic and diastolic heart failure | I50.40 | unspecified combined systolic (congestive) and diastolic (congestive) heart failure |
| 428.40 | combined systolic and diastolic heart failure, unspecified | I50.41 | acute combined systolic (congestive) and diastolic (congestive) heart failure |
| 428.41 | acute combined systolic and diastolic heart failure | I50.42 | chronic combined systolic (congestive) and diastolic (congestive) heart failure |
| 428.42 | chronic combined systolic and diastolic heart failure | I50.43 | acute on chronic combined systolic (congestive) and diastolic (congestive) heart failure |
| 428.43 | acute on chronic combined systolic and diastolic heart failure | I50.8 | other heart failure |
| 428.9 | heart failure, unspecified | I50.81 | right heart failure |
| 402.01 | malignant hypertensive heart disease with heart failure | I50.810 | right heart failure unspecified |
| 402.11 | benign hypertensive heart disease with heart failure | I50.811 | acute right heart failure |
| 402.91 | unspecified hypertensive heart disease with heart failure | I50.812 | chronic right heart failure |
| 404.01 | hypertensive heart and chronic kidney disease, malignant, with heart failure and with chronic kidney disease stage I through stage IV, or unspecified | I50.813 | acute on chronic right heart failure |
| 404.03 | hypertensive heart and chronic kidney disease, malignant, with heart failure and with chronic kidney disease stage V or end stage renal disease | I50.814 | right heart failure due to left heart failure |
| 404.11 | hypertensive heart and chronic kidney disease, benign, with heart failure and with chronic kidney disease stage I through stage IV, or unspecified | I50.82 | biventricular heart failure |
| 404.13 | hypertensive heart and chronic kidney disease, benign, with heart failure and chronic kidney disease stage V or end stage renal disease | I50.83 | high output heart failure |
| 404.91 | hypertensive heart and chronic kidney disease, unspecified, with heart failure and with chronic kidney disease stage I through stage IV, or unspecified | I50.84 | end stage heart failure |
| 404.93 | hypertensive heart and chronic kidney disease, unspecified, with heart failure and chronic kidney disease stage V or end stage renal disease | I50.89 | other heart failure |
| | | I50.9 | heart failure, unspecified |

*Data used for prediction*: We used all structured EHR data, including patient demographics, diagnoses, medications, and procedures occurred before the HFOD for prediction. Clinical variables occurred at or after HFOD were not used.

*Case-control matching*: For each case, up to 9 controls were selected according to criteria defined as follows: (1) having the same gender and race as the case; (2) age is within five-year interval of the case; (3) having the same major cancer type; (4) the first encounter date of the control is within a year with the first encounter date of the case; (5) having an encounter occurred either within 30 days before or any time after the HFOD of the case. The encounter date closest to the HFOD is denoted as reference encounter date. For cases that could not be matched to any controls, we removed them from the data set. To facilitate comparison with previous studies, we followed the same case-control matching procedure without replacement. Figure 1 shows an overview of the case-control matching procedure.

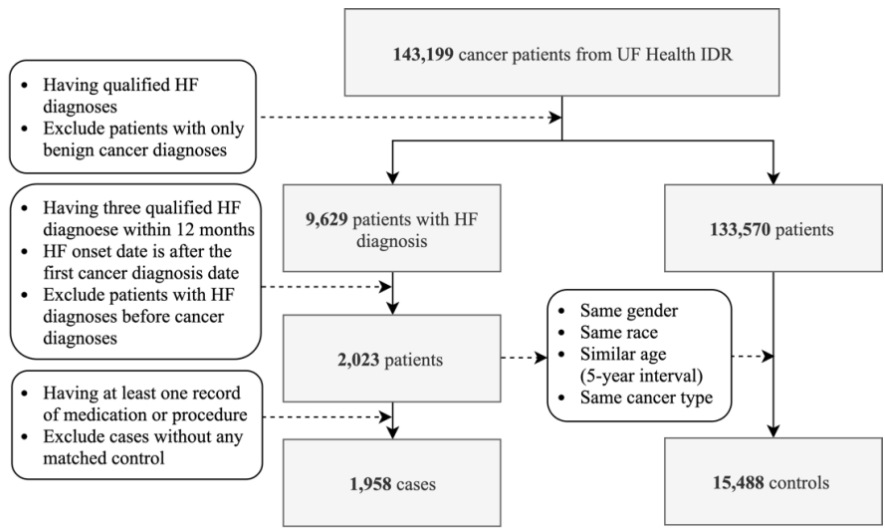

**Figure 1**. An overview of case-control matching procedure

**Machine learning algorithms**

In this study, we explored four widely used machine learning algorithms including LR, SVMs, RFs, and GB. We applied LR as the baseline method and compared it with SVMs, RFs, and GB. For LR, RFs, and SVMs, we adopted implementations in the sciki-learn library (http://scikit-learn.org). For GB, we used the implementation from the XGBoost package (https://github.com/dmlc/xgboost)[22]. For LR, we optimized the optimization method (solver), the regularization parameter *c*, and the tolerance of termination criterion *e*. For SVMs, we used the Radial Basis Function (RBF) kernel and tuned the regularization parameter *c*, and the tolerance of termination criterion *e*. For RFs, we performed the optimization on the parameters including the number of trees (n_estimators), the tree splitting function, the tree max depth (max_depth). For GB, we optimized the learning rate (eta), the maximum depth of a tree (max_depth), and the number of boost trees (n_estimators) and chose the logistic regression for binary classification as the objective function to evaluate the training loss and regularization.

**Variable grouping and feature encoding**

In this study, we used patient demographics (gender, race, age), diagnoses, medications, and procedures to develop machine learning models for heart failure prediction. As previous studies[19,21] have identified the sparseness issue of using the diagnoses and medications as features, we grouped the diagnoses codes (ICD-9 or ICD-10) and medication codes (RxNorm concept unique identifier codes or drug names) according to existing disease groupings.

For diagnoses codes, one previous study from Choi *et al.* grouped the ICD 9 codes using the Clinical Classification Software (CCS)[23] groups (having a total number of 283 groups) and Rasmy *et al.* grouped ICD 9 codes using PheWAS (Phenome-wide association studies) groups (about 1,800 unique PheWAS groups). Following the study of Rasmy *et al.*, we mapped the ICD-9 and ICD-10 codes to the PheWAS groups. [24,25] Using the PheWAS code translation table[24], we mapped a total number of 36,701 unique ICD-9 and ICD-10 codes into a total of 4,973 unique PheWAS groups.

For medication codes in RxNorm concept unique identifiers (RXCUIs), we mapped them to the ingredient-level RXCUIs. For example, Abraxane (with RXCUI 589511) is a brand name for a chemotherapy drug with the ingredient

Paclitaxel (RXCUI 56946). Thus, we mapped all RXCUI of 589511 to RXCUI 56946 when developing machine learning models. After mapping all medications to their ingredient level RXCUI codes, we mapped a total number of 19,774 clinical-level RXCUI codes to a total of 2,376 ingredient-level RXCUI codes.

To convert the clinical variables into machine-readable features for training, we compared two encoding methods. The first method is one-hot encoding where variables are encoded using binary values of '1' (indicating feature occurred in a sample) or 0 (indicating feature not occurred). The resulted one-hot encoded data is a sparse matrix of 1 and 0. In the one-hot encoding, the frequency of variables is not considered. Variables occurred multiple times were counted only once. In the second encoding method, we further considered the frequency of variables among patients. Inspired by the term frequency-inverse document frequency (TF-IDF) strategy[26], we treat each patient as a document and each feature as a word in the document. Then, we applied the standard TF-IDF calculation to convert each patient into a TF-IDF weighted vector. For each machine learning algorithm, we trained two models for one-hot encoding and TF-IDF encoding, respectively.

**Experiments and evaluation**

Using stratified sampling, we split the data into a training set with 13,956 patients (1,566 cases and 12,390 controls) and a test set with 3,490 patients (392 cases and 3,098 controls). We optimized the machine learning models using five-fold cross validation and grid searching. We trained machine learning models using the training set and evaluated the performances using the test set. Following the previous studies, we used the area under the receiver operating characteristic curve (AUC or AUC-ROC) for evaluation. We also conducted statistical tests to compare different methods. The statistical test scores were calculated by sampling 100 times from the test data and each time a number of 1,745 (50% of the test set) samples were randomly selected. We used the t-test to calculate p-values. We also reported the sensitivity and specificity determined using the Youden's index to facilitate comparison.[27]

**Results**

**Table 2.** Comparison of case group and control group.

| Variables | Sub Types | Case (n=1,958) | Control (n=15,488) | *p*-value |
|---|---|---|---|---|
| **Diagnoses** | | 14.3[a] (2.1[b]) | 12.5 (2.5) | < 0.001[c] |
| **Medications** | | 87.7 (47.3) | 35.9 (35.4) | < 0.001[c] |
| **Procedures** | | 20.4 (5.2) | 16.2 (4.3) | < 0.001[c] |
| **Gender** | Male | 1,067 (54.5%) | 8,518 (55.0%) | 0.691[d] |
| | Female | 891 (45.5%) | 6,970 (45.0%) | |
| **Age** | <40 | 57 (2.9%) | 361 (2.3%) | 0.007[d] |
| | 40-44 | 22 (1.1%) | 147 (0.9%) | |
| | 45-49 | 47 (2.4%) | 299 (1.9%) | |
| | 50-54 | 84 (4.3%) | 685 (4.4%) | |
| | 55-59 | 169 (8.6%) | 1,260 (8.1%) | |
| | 60-64 | 197 (10.1%) | 1,739 (11.2%) | |
| | 65-69 | 263 (13.4%) | 2,394 (15.5%) | |
| | 70-74 | 335 (17.1%) | 2,850 (18.4%) | |
| | 75-79 | 275 (14.0%) | 2,473 (16.0%) | |
| | 80-84 | 249 (12.7%) | 1,786 (11.5%) | |
| | ≥85 | 260 (13.3%) | 1,494 (9.6%) | |
| **Race** | White | 1,578 (80.6%) | 12,849 (83.0%) | < 0.001[d] |
| | African American | 304 (15.5%) | 1,359 (8.8%) | |
| | Asian | 8 (0.4%) | 116 (0.7%) | |
| | Other | 69 (3.5%) | 1,164 (7.5%) | |

[a] mean value; [b] standard deviation; [c] derived from the *t*-test; [d] derived using the chi-squared contingency test

After case-control matching, we identified a total number of 1,958 cases and 15,488 controls. Table 2 compares the descriptive statistics between the case and control groups. For demographic information, the case and control groups have a similar distribution as a result of the matching procedure. On average, patients in the case group have more diagnoses, medications, and procedures compared to the control group patients. In both case and control groups, the most common cancer subgroups include cancer of skin, female breast cancer, prostate cancer, and lung cancer.

Table 3 compares the performance of four machine learning models using two different encoding strategies. The last column shows the statistical p-values between two encoding strategies for each method. The baseline LR achieved an AUC of 0.8795 and 0.8762 for one-hot encoding and TF-IDF encoding, respectively. The statistical test (p-value of 0.1072) showed that there is no significant difference between the two encoding strategies for LR. For other machine learning methods, the TF-IDF encoding outperformed the One-hot encoding with significant p-values. The GB model trained with TF-IDF encoding achieved the best AUC of 0.9077 outperforming all other models with statistical p-values of <0.01. The AUC scores of the SVMs-based models are notably lower than other machine learning models (p-values of < 0.001), which is consistent with the experimental results reported by Wu et al.[18] in a similar study. The GB model trained with one-hot encoding obtained the best sensitivity of 0.8546 and the SVMs trained with one-hot encoding obtained the best specificity of 0.8657.

**Table 3.** Comparison of machine learning methods using one-hot encoding and TF-IDF encoding.

| Algorithm | Data Encoding Method | Sensitivity | Specificity | AUC | p-value |
|---|---|---|---|---|---|
| LR | One-hot encoding | 0.8418 | 0.7692 | 0.8795 | 0.1072 |
| LR | TF-IDF encoding | 0.7449 | 0.8589 | 0.8762 | |
| SVMs | One-hot encoding | 0.7423 | **0.8657** | 0.8473 | <0.0001 |
| SVMs | TF-IDF encoding | 0.7347 | 0.8518 | 0.8314 | |
| RFs | One-hot encoding | 0.7730 | 0.8325 | 0.8737 | <0.0001 |
| RFs | TF-IDF encoding | 0.8240 | 0.7824 | 0.8884 | |
| GB | One-hot encoding | **0.8546** | 0.7795 | 0.8938 | 0.0016 |
| GB | TF-IDF encoding | 0.8520 | 0.8138 | **0.9077** | |

- LR: logistic regression; SVMs: support vector machines; RFs: random forests; GB: gradient boosting; TF-IDF: term frequency-inverse document frequency
- Best AUC, sensitivity, and specificity are highlighted in bold
- p-values were used to compare the performances of the two encoding methods for each machine learning algorithm.

**Discussion and Conclusion**

In this study, we examined four machine learning algorithms for heart failure prediction using cancer patients' EHR data from UF Health IDR. Starting from a total number of 143,199 cancer patients, we identified 1,958 qualified cases who developed heart failure after diagnoses of cancers, which were matched to 15,488 controls by gender, age, race, and major cancer type. We compared two feature encoding strategies including one-hot encoding and TF-IDF encoding in developing machine learning methods. The GB model with TF-IDF feature encoding achieved the best AUC score of 0.9077, significantly outperforming other machine learning methods. The experimental results show that it is feasible to use machine learning methods to identify cancer patients with risks of cardiotoxicity. Previous studies have applied machine learning methods for heart failure prediction among general patient cohorts without anchoring on any diseases. This study demonstrated the feasibility of using machine learning methods for heart failure prediction among cancer patients. Among the four machine learning models, LR and RFs achieved decent performance. However, the performance of the SVMs model is relatively lower than other machine learning methods. A similar study by Wu et al.[18] also reported that SVMs achieved lower performance in general patient cohorts. We also explored two feature encoding strategies including one-hot encoding and TF-IDF encoding. The experimental results (Table 3) showed that the models trained with the TF-IDF encoding method achieved better performance compared to one-hot encoding method for SVMs, RFs, and GB. A possible reason is that the TF-IDF encoding can capture frequency information of variables thus to enhance performance.

We looked into the subgroup of cancer patients with exposure to chemotherapy drugs to further examine the prediction performance. To facilitate subgroup analysis, two UF Health physicians (NW and KM) manually reviewed a total number of 1,557 ingredient-level medications used by at least 10 patients and identified 104 chemotherapy drugs. The top three drugs are methotrexate (used by 1,486 patients), cyclophosphamide (used by 1,371 patients), and carboplatin (used by 1,275 patients). Using this list of chemotherapy drugs, we identified 458 patients with exposure to chemotherapy drugs from the test set and calculated the sensitivity and specificity. For this subgroup, the best machine learning model (i.e., GB) achieved a sensitivity of 0.8824 and a specificity of 0.6300, respectively. Compared with the scores on the entire test set (a sensitivity of 0.8520 and a specificity of 0.8138), this subgroup has a better sensitivity but a significantly lower specificity. This may indicate that exposure to chemotherapy drugs is a strong but not deterministic risk factor for heart failure detection. As many chemotherapy drugs are known to have cardiotoxic side effects, it's not surprising to see that chemotherapy drugs are strong risk factors. Yet, the low specificity indicates that it is still challenging for machine learning models to figure out why some cancer patients develop cardiotoxicity after exposure to chemotherapy treatments while others do not.

This study focused on the prediction of cancer patients who developed heart failure after cancer diagnoses. Compared with the results from general patient cohorts, the machine learning models achieved a better AUC score in our cancer patient cohort constructed from the UF Health IDR. We compared the statistics between our cancer patient cohort with general patient cohorts used in previous studies. The cancer patient cohort used in this study has more rich medication and clinical procedure information. For example, the average numbers of medications and clinical procedures in a general patient cohort constructed by Rasmy *et al.* from the Cerner Healthfacts® dataset[21] are remarkably lower (27.26 in cases and 1.4 in controls) than our cancer patient cohort (87.7 in cases and 20.4 in controls). This may be one of the reasons that the machine learning methods achieved a better AUC on our cancer patient cohort.

We also calculated the feature importance using the best machine learning model (i.e., GB with TF-IDF encoding method). The results show that diagnosis of essential hypertension, atrial fibrillation, shortness of breath, tobacco use disorder, and hyperlipidemia are the top risk factors for heart failure prediction. These factors are clinically known risk factors for heart failure, indicating that the machine learning models are able to capture important clinical factors for heart failure prediction. Our ultimate goal is to develop predictive models to identify cancer patients with high risks of cardiotoxicity to prevent or minimize the risk of cardiotoxicity in cancer treatments. This study demonstrated the feasibility of machine learning methods for the prediction of cancer patients developing heart failure after cancer diagnoses. Even though machine learning methods can identify important clinical factors, the mechanism of cancer therapy-induced cardiotoxicity is largely unknown. The low specificity score on the subgroup of cancer patients with exposure to chemotherapy drugs shows that it's still challenging for machine learning models to identify cancer patients with high-risk of cardiotoxicity after chemotherapy. This study is a preliminary step to assess machine learning models for heart failure prediction among cancer patients. To further improve the performance for cancer patients with exposure to chemotherapy drugs, we plan to explore advanced machine learning models such as deep learning and non-clinical factors such as patient genomics data.

This study has limitations. Similar to other EHR-based studies, our study may suffer from incomplete information, varying length of observation, and coding bias. We carefully designed the cancer patients who later developed into heart failure, yet, some of the cases may not necessarily come from cancer treatments. We expect more carefully designed studies to further evaluate our findings.

**Acknowledgement**

This study was supported by the University of Florida Clinical and Translational Science Institute, which is supported in part by the NIH National Center for Advancing Translational Sciences under award number UL1TR001427. The content is solely the responsibility of the authors and does not necessarily represent the official views of the National Institutes of Health.